\newcommand{\defeq}[1]{\begin{equation}\label{#1}}
\newcommand{\eeq}{\end{equation}}
\documentclass{article}

\usepackage{amsfonts}
\usepackage{amsmath}

\usepackage{amssymb}
\usepackage{amsthm}

\usepackage{framed}
\usepackage{algorithm}
\usepackage{algorithmicx}
\usepackage{algpseudocode}

\usepackage[margin=1.0in]{geometry}
\usepackage{multicol}
\usepackage{multirow}
\usepackage{bigdelim}

\usepackage{etoolbox}

\usepackage{color} 
\definecolor{shadecolor}{gray}{0.9}
\definecolor{lightyellow}{RGB}{255,255,192}
\definecolor{lightblue}{RGB}{192,192,255}
\definecolor{lightpurple}{RGB}{255,192,255}
\definecolor{lightgreen}{RGB}{192,255,192}
\usepackage{mdframed}

\usepackage{array}
\usepackage{colortbl}
	\definecolor{CellColor}{RGB}{255,255,192}
	\definecolor{HighlightColor}{RGB}{255,192,192}
	\definecolor{HighlightColorB}{RGB}{192,255,192}

	\newcolumntype{C}{>{\columncolor[RGB]{255,255,192}}c}
	\newcolumntype{L}{>{\columncolor[RGB]{255,255,192}}l}
	\newcolumntype{R}{>{\columncolor[RGB]{255,255,192}}r}
	\arrayrulecolor[RGB]{0,0,0}
	\doublerulesepcolor[RGB]{0,0,0}
\usepackage{hhline}

\usepackage{float}
\usepackage{caption}[2007/09/01] 

\usepackage{tikz}
	\usetikzlibrary{arrows}
	\usetikzlibrary{decorations.fractals}
	\usetikzlibrary{decorations.pathreplacing}
	\usetikzlibrary{patterns}
	\usetikzlibrary{shapes}

\usepackage{outlines}
\usepackage{enumitem}
	\setenumerate[4]{label=\alph*.}
	\setenumerate[1]{label=\Roman*.}
	\setenumerate[2]{label=\Alph*.}
	\setenumerate[3]{label=\roman*.}

\usepackage{url}
\usepackage{hyperref}		

\numberwithin{equation}{section}	

\theoremstyle{plain}			

\theoremstyle{definition}		

\renewcommand{\qed}{\hfill \mbox{\raggedright \rule{.07in}{.1in}}}

\DeclareCaptionLabelFormat{underlcap}{\bf{$\underline{\text{Figure #2}}$}}
\newenvironment{ColumnFigure}
{
	\begin{table}
	\begin{framed}
}
{
	\end{framed}
	\end{table}
}

\newcommand\Section[2][]{\begin{mdframed}[skipabove=12pt,skipbelow=0.5cm, leftline=false, rightline=false, bottomline=true, topline=false, linecolor=black, linewidth=2,innertopmargin=10pt,innerbottommargin=10pt]   \ifx\relax#1\relax\section{#2}\else\section[#1]{#2}\fi   \end{mdframed}} 

\newcommand\Subsection[2][]{\begin{mdframed}[skipabove=12pt,skipbelow=0.5cm, leftline=false, rightline=false, bottomline=true, topline=false, linecolor=black, linewidth=2,innertopmargin=10pt,innerbottommargin=10pt]   \ifx\relax#1\relax\subsection{#2}\else\subsection[#1]{#2}\fi   \end{mdframed}} 

\patchcmd{\thebibliography}{\section*}{\Section}{}{}

\newcommand{\TitleString}{The Polynomial Transform} 
\title{\TitleString\\
\copyright \small{Matt Groff 2019} \\
} 
\author{MATT GROFF\\
P.O. Box 642\\
Camp Hill, PA, USA 17001-0642\\
mgroff100@hotmail.com}

\begin{document}
\pagestyle{myheadings}

\markright{\rlap{\TitleString}\hfill \textnormal{\copyright Matt Groff 2019}\hfill}


\twocolumn[\maketitle]


\begin{abstract}
We explore a new form of DFT, which we call the Polynomial Transform.  It functions over finite fields, and a size $n$ transform takes $O(n)$ arithmetic operations.  In the multitape Turing machine model, it allows us to multiply two $n$ bit numbers in time $n \left( k^{ \log^*{n} } + \log{p} \right)$, where $k$ is a constant and $\log^*{n}$ is the iterated logarithm.  One important consequence is that the Network Coding Conjecture is false.
\end{abstract}

\Section{Introduction}
Our algorithm, like many of the multiplication algorithms before, relies on the DFT, the DFT is the major bottleneck in multiplication algorithms due to its' $O(n \log{n})$ time.


The DFT is a very useful algorithm.  With the popularization of the FFT by Cooley and Tukey in 1965\cite{Cooley-Tukey-DFT}, it became much more widely used.  Since then, there have been a few attempts at speeding up the DFT.  In 2012 there was a major breakthrough with Hassanieh, Indyk, and Price's sparse FFT (sFFT)\cite{arxiv:sFFT}, that gave a time of $O(k \log{n})$ for exactly $k$-sparse algorithms and $O(k \log{n} \log{n/k})$ for the general case.  One drawback of the algorithm is that it needs $n$ to be a power of 2.

In this paper, we present a DFT algorithm that uses only in $O(n)$ operations.  Like some forms of the DFT, it only works for certain sizes, although it is much less limited than many.  Then we use this to multiply two naturals in $N K^{ \log^*{N} }$ time.  

We note that it was Karatsuba, who in 1962, first improved on the na{\"{i}}ve bound of $n^2$ to ${n^{\log_2{3}}}$.

Besides the straight line program asymptotic limit \cite{Straight-Line-Bounds}, Sch{\"o}nhage and Strassen conjectured in 1971 that integer multiplication must take time $\Theta{(n \log{n})}$\cite{Schonhage-Strassen}.  In that same paper, they set the world record at $O(n \log{n} \log{\log{n}})$.  That bound stood for almost 40 years, until F{\"u}rer got the new bounds $O \left( n \log{n}K^{\log^*{n}} \right)$, where $K$ is some constant and $\log^*{}$ is the iterated logarithm\cite{Furer-paper}.  Finally, there were a number of results, particularly with a number of impressive papers from Harvey and van der Hoeven, that culminated in their $O(n \log{n})$ time algorithm\cite{Harvey-van-der-Hoeven}.

\Section{Polynomial Transform Overview}
We use the multitape Turing machine as our model of computation.

We start with an easy and familiar transform, the DFT over the complex integers, for which we will really use only natural numbers.  Now, we start assuming that we want a transform of size $n$, and will perform the transform modulo a prime $p$, with $n = p^2-1$.  This means that we have to find an $n$th \textit{principal} root of unity, $\omega$. 


Now, the trick of the algorithm is that we return the result of a ``transform'' modulo $p$, but do not perform the calculations modulo $p$.  We perform it modulo slightly smaller primes from the set $\{q_1, q_2, \dots, q_k, \dots, q_\alpha\}$.  We pick a prime $p$ such that it does not have an element equivalent to the imaginary number $i$, in some senses.  In other words, modulo $p$, there is no natural number $z$ such that $z^2 \equiv -1$.  Also, we ensure that all other primes $q_k$, that we use, do ``have'' an imaginary number.  This ensures that each $q_k$ actually has $2$ imaginary numbers, since having one square $z^2 \equiv -1 \mod q_k$ ensures that another square is also equivalent to $-1$.  We can call them $i_1$ and $i_2$.

The real part of the trick is now that we can evaluate the DFT, modulo $q_k$, using both $i_1$ and $i_2$ as the imaginary number $i$.  When we evaluate the DFT modulo the smaller primes $q_k$, all of our outputs are in the form of a complex number $z_r + z_i i$, where $z_r$ is the real component and $z_i$ is the imaginary component.  We can evaluate the DFT twice modulo $q_k$, for which the first outputs are then in the form $z_r + z_i i_1$, and the second outputs are in the form $z_r + z_i i_2$.  We can now use linear algebra on the results to recover $z_r$ and $z_i$ from the results.  This allows us to evaluate the DFT using only a $O(\sqrt{n})$ sized DFT, as opposed to a size $n$ DFT, and it will give us the correct results for the $\ell$th output $\widehat{x_\ell}$ as $(\widehat{x_\ell} \mod p) \mod q_k$.

Now, to recover the $\ell$th output $(\widehat{x_\ell} \mod p$ from the outputs $(\widehat{x_\ell} \mod p) \mod q_k$, we simply use the Chinese remainder theorem on the various moduli $q_k$ from $\{q_1, q_2, \dots, q_k, \dots, q_\alpha\}$.  Assuming that all of our calculations use somewhat small numbers, we shouldn't have to use more than a constant number of primes.  We'll make this more rigorous shortly.

\Subsection{A Quick Example}
We can demonstrate this component reduction with complex numbers.  Modulo $5$, we can take the imaginary number $i$, and square it to get $i^2 = -1$.  And $-1 \equiv 4 \bmod 5$.  But $2^2 \equiv 4 \bmod 5$, and $3^2 = 9 \equiv 4 \bmod 5$.  So any complex number, modulo $5$ \textit{reduces} to $a + bi \equiv a + 2b \bmod 5$ or $a + bi \equiv a + 3b \bmod 5$.  The idea is then to use both versions, and use linear algebra to find the components modulo $5$.

An example is $3 + 1 i \bmod 5$.  First, setting $i \equiv 2 \bmod 5$ yields $3 + 1(2) = 5 \equiv 0 \bmod 5$.  Then, setting $i \equiv 3 \bmod 5$ yields $3 + 1(3) = 6 \equiv 1 \bmod 5$.  Now we use linear algebra (modulo $5$) to recover the real and imaginary components:

\begin{align}
	\begin{bmatrix}
		3(1) = 3 & 1(2) = 2 \\
		3(1) = 3 & 1(3) = 3
	\end{bmatrix} 
	\begin{bmatrix}
		1 \\
		i
	\end{bmatrix} =
	\begin{bmatrix}
		0 \\
		1
	\end{bmatrix}
\end{align}

From this matrix we recover real coefficient $3$ and imaginary coefficient $i = 1$.

\Subsection{Algorithm Overview \label{Algorithm Overview Section}}
To go into more detail, and to explain the abstract algebra we use, we really start with doing operations modulo $p$, and adjoin the cube root $\sqrt[3]{y}$ and its' square $\left( \sqrt[3]{y} \right)^2$, for some $y$ we'll choose later.  We then use the two roots $\sqrt[3]{y}$ and $\left( \sqrt[3]{y} \right)^2$ similarly to how the imaginary number $i$ is used with the complexes.  Then, we pick primes $q_k$, for which $\sqrt[3]{y}$ and $\left( \sqrt[3]{y} \right)^2$ evaluate to naturals.  For example, if we use calculations modulo $q_k = 11$, then we could use $y=4$ since $4^3 = 9$, and therefor $\sqrt[3]{9} \equiv 4 \mod 11$\footnote{This is very much like working with $\mathbb{F}(x) / r(x)$, for $\mathbb{F} = \mathbb{Z}/n \mathbb{Z}$.}.

Since we find a few different $q_k$ values to use, we can use the Chinese remainder on the coefficients modulo $q_k$, to recover them modulo $m$ if we are very careful.  The idea is that a $3$ component system will have a maximum value, from one multiplication modulo $p$, of $3 p^2$ for each coefficient.  This comes from the fact that each coefficient is at most $p$, so that when we multiply 2 coefficients together, we get at most $p^2$ as the value, and since there are $3$ different coefficients, we multiply each coefficient by 3.  We are not done, because to convert these 3 components into the reduced single value modulo $q_k$, we must multiply each again by at most $p$ and then sum the coefficients, to get a total value of size $3(p(3p^2)) = 9p^3$.  That gives a single coefficient.  Now, to be safe, we want our system to handle all of the coefficients summed into one value, so this means that we have $p^3$ coefficients of size at most $9p^3$ summed together into one coefficient modulo $q_k$, which means that our coefficients are at most $9p^6$.

We can use the Chinese remainder theorem to ensure that we can recover our $3$ coefficients, if the primes, when combined together, can handle an exact value of at most $9p^6$.  In other words, we want $\prod_k{ q_k } > 9p^6$.

We note that the order of the roots that we use do not matter when we combine our matrices, since the values of the roots are changed, but will still give the same result.  However, we know which root is $\sqrt[3]{y}$ and which is $\left( \sqrt[3]{y} \right)$, so that we technically order them.

\begin{algorithm*}
	\begin{algorithmic}[1]
		\Procedure{PreprocessPT}{ $N$ }
			\State Find $p$
			\State Find a set of usable primes $\{q_1, q_2, q_3, \dots, q_\alpha \}$
			\State Determine $\omega$, a root of unity modulo (a field extension of) $p$ of order $p^3$
			\State Determine coefficients' multipliers $\omega^{j k}$ of DFT/Vandermonde matrix modulo each prime $q_k$
		\EndProcedure
		
		\Procedure{PT}{ $\{X\}$ }
			\State \Call{PreprocessPT}{ $| \{X\} |$ }
			\ForAll{$q_k$}
				\State Calculate coefficient list
				\State Perform DFT modulo $q_k$
			\EndFor
			\State \Return $\{ \widehat{X} \}$
		\EndProcedure
	\end{algorithmic}
	\caption{Polynomial Transform Algorithm\label{Polynomial Transform Algorithm}}
\end{algorithm*}

\Section{Running Time Analysis}
A sketch of the algorithm is shown in Figure \ref{Polynomial Transform Algorithm}.

We can examine the preprocessing first.  Our first task is to find the prime $p$, for which $n \approx p^3$.  For this we can use the sieve of Atkin and Bernstein\cite{SieveOfAtkin}, which finds all primes less than $p$.  It takes $O(p)$ arithmetic operations, for a running time of $O(p \log{p})$, assuming that each operation takes time $O(\log{p})$, since we don't use any numbers larger than $O(\log{p})$ bits for this task.  So we can certainly find a prime large enough to use for $p$ in time linear in $n$, but the problem then becomes what $p$ will be, exactly.

We can certainly look at the formula for what we want $p$ to be, exactly.  We want $p^3 \approx n$ for the DFT.  If we start by taking an estimate $p_0$ for $p$, and repeatedly doubling it, we will reach $\sqrt[3]{n}$ very quickly, in $O(\log{n})$ operations.  So that is what we do.  We keep track of our estimates for $N$, which we can call $n_0$, and our estimate for $p$, $p_0$.  So we double $p_0$ each iteration, and to get our next $n_0$, we take:

\begin{align}
	n &< (2p)^3 \\
	  &< 8 p^3
\end{align}
	
So we successively multiply our approximation of $n$ by $8$, every time we double our approximation for $p$.  These can both be done easily by bit shifts, so we can certainly find a minimum value for $p$ fairly quickly.  Then we just use the next largest prime for this $p$.


After we have found $p$, we pick one value for $\sqrt[3]{y}$,and cube it.  This gives us $y$ and our cube root $\sqrt[3]{y}$, and this also gives us $\left( \sqrt[3]{y} \right)^2$.  We have to ensure that this root does not exist in $\mathbb{Z}/p \mathbb{Z}$ already.  So we cycle through values until we find one that does not exist already.  This happens $2/3$ of the time for each choice of $y$ with $p \equiv 2 \mod 3$, due to the \textit{Chebotarev theorem}, explained in \cite{Chebotarev}.

  In fact, we must ensure that $m \equiv 2 \mod 3$, because $m$ is not a prime when $m \equiv 0 \mod 3$, and if  $m \equiv 1 \mod 3$, then every element in this field has a cube root.  We can see this through Fermat's little theorem.  We have

\begin{align}
	y^{q_k}            &\equiv x \mod q_k \\
	y^{q_k - 1}        &\equiv 1 \mod q_k \\
	y^{q_k}y^{q_k - 1} &\equiv x \mod q_k \\
	y^{2q_k - 1}       &\equiv x \mod q_k
\end{align}

Where the first two equations are Fermat's little theorem, and the last two follow from multiplying each side of the first two together.  If we then substitue $3r+2$ for $q_k$, we get

\begin{align}
	y^{2q_k - 1}       &\equiv x \mod q_k \\
	y^{2(3r+2) - 1}    &\equiv x \mod q_k \\
	y^{6r+3}           &\equiv x \mod q_k
\end{align}

This is taken from \cite{website:Wikipedia-cubic-reciprocity}.

\begin{ColumnFigure}
	\begin{center}
		$\begin{bmatrix}
			c_1         & c_2     & c_3    & \ddots  & c_{q_k} \\
			c_{q_k}     & c_1     & c_2    & \ddots  & c_{q_k-1} \\
			c_{q_k - 1} & c_{q_k} & c_1    & \ddots  & c_{q_k-2} \\
			\ddots      & \ddots  & \ddots & \ddots & \vdots \\
			c_2         & c_3     & c_4    & \dots  & c_1 \\
		\end{bmatrix}$
	\end{center}
	\caption{\bf{A Circulant matrix}\label{Circulant Matrix Figure}}
\end{ColumnFigure}

We can also find the various cube roots by cubing each value from $0$ to $q_k - 1$ modulo $q_k$ which takes 2 multiplications per value, and thus $O(q_k \log{q_k})$ time per prime $q_k$.  Since we're only interested in storing the two cube roots of $y$, this takes $O(\log{q_k})$ space.  Here, we simply want to find two cube roots that do not exist modulo $p$.  They should both exist modulo all primes $q_k$ that we use, other than $p$, so that we can use linear algebra on them (as described above).  All other primes that we find, we can start with primes close to $p$ and proceed towards smaller and smaller primes.  That the probabilities that values modulo some prime $q_k$ are cubes is roughly equal, according to \cite{Random-Cubes}.  This ensures us that we can find a cube root $\sqrt[3]{y}$ with probability roughly $1/3$.  Also, if we use a circulant matrix, which has the form given in Figure \ref{Circulant Matrix Figure}, where all diagonal entries are the same, we have the the probability that the $3 \times 3$ matrix created from the corresponding cube roots is singular is approximately $1/{q_k}$.  Thus we should be able to find a set of usable primes fairly easily, since the number of primes is constant.  This follows because they are all fairly close to $p$; that is to say $q_k \approx p$.  Also, the circulant matrix, taken from \cite{Circulant-Determinant} is known to have determinant

\defeq{Circulant Equation}
	\prod_{j=0}^{q_k - 1}{ \sum_{l=0}^{q_k - 1}{ c_k \omega^{j k} } } \mod q_k \text{ where }\omega = e^{2 \pi i / (q_k-1)}
\eeq

This equation will equal $0$ modulo $q_k$ approximately $1/{q_k}$ of the time, since the entries or coefficients of the matrix are again evenly distributed.  Thus as $p$ goes to infinity, so do all of the potential $q_k$ values that we use, and thus the probability becomes smaller that the matrices are singular.  Namely, it approaches $1/3$.


Now, to find the set of usable primes, we should know that this is equivalent to finding an integer that functions as the cube root of $y$ modulo $q_k$, and as we saw above, this takes time $O(q_k \log{q_k})$ time per prime.  Now, the asymptotic formula for this, according to \cite{Cohen-Prime-Bound}, is

\defeq{p log p Equation}
	\sum_{\substack{ q_k \text{ Prime} \\ q_k < m}}{ q_k \log{q_k} } = O(p^2 \text{ln}(p)^2)
\eeq

This is certainly done within $O(p^3) \approx O(N)$ time.

To find $\omega$, we can use the results described in most standard textbooks, for instance see \cite{ANTv1}.

Finally, we will address smaller DFTs here and in the next subsection.  For each prime $q_k$, we compute each individual power of $\omega^{j k}$ to associate with the input coefficients and rows and columns of each DFT matrix.  This takes $O(q_k)$ operations, which is certainly in $O(p^3)$, as each prime $q_k$ that we use will be much less than $p$.

\begin{ColumnFigure}
	\begin{center}
		$\begin{bmatrix}
			\left( v_1 \right)^{q_k} & \left( v_2 \right)^{q_k} & \left( v_3 \right)^{q_k} & \dots  & \left( v_{q_k} \right)^{q_k} \\
			\left( v_1 \right)^1     & \left( v_2 \right)^1     & \left( v_3 \right)^1     & \dots  & \left( v_{q_k} \right)^1 \\
			\left( v_1 \right)^2     & \left( v_2 \right)^2     & \left( v_3 \right)^2     & \dots  & \left( v_{q_k} \right)^2 \\
			\left( v_1 \right)^3     & \left( v_2 \right)^3     & \left( v_3 \right)^3     & \dots  & \left( v_{q_k} \right)^3 \\
			\vdots                   & \vdots                   & \vdots                   & \ddots & \vdots \\
		\end{bmatrix}$
	\end{center}
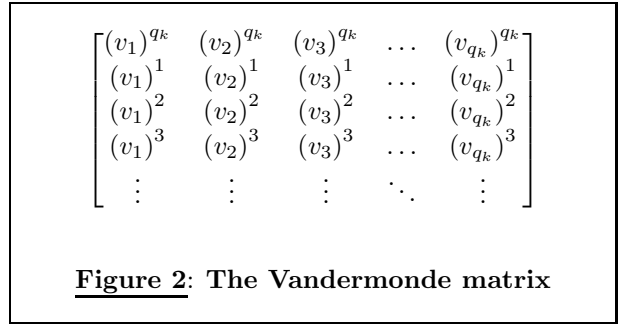
	\caption{\bf{The Vandermonde matrix}\label{Vandermonde Matrix Figure}}
\end{ColumnFigure}

\begin{ColumnFigure}
	\begin{center}
		$\begin{bmatrix}
			[V]    & [V]    & [V]    & \dots \\
			[V]    & [V]    & [V]    & \dots \\
			[V]    & [V]    & [V]    & \dots \\
			\vdots & \vdots & \vdots & \ddots \\
		\end{bmatrix}$
	\end{center}
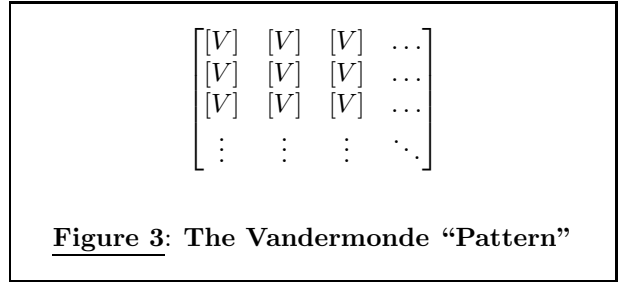
	\caption{\bf{The Vandermonde ``Pattern''}\label{Vandermonde Pattern Figure}}
\end{ColumnFigure}

\Subsection{The algorithm \label{Algorithm Section}}
The first step of the algorithm is to collect all of the coefficients modulo each prime $q_k$.  What we mean by this is in Figure \ref{Vandermonde Matrix Figure} and Figure \ref{Vandermonde Pattern Figure}.  Performing the DFT is equivalent to multiplying a Vandermonde matrix by a vector.  This equivalent Vandermonde matrix is the matrix in Figure \ref{Vandermonde Matrix Figure}, with $V_k$ replaced by $\omega^{k-1}$, where $\omega$ is an element with multiplicative order $n$, for an $n$ element DFT.

Now, if we take our $p^3 > n$ element Polynomial Transform, which is essentially like an $p^3$ element DFT, then the matrix equivalent of the Polynomial transform, modulo $q_k$, is the matrix in Figure \ref{Vandermonde Pattern Figure}.  Therefor, modulo our primes $q_k$, for all $k$, this matrix becomes a Vandermonde matrix.  Here we can get the correct Vandermonde matrix by adding together the coefficients that are matched with each element $V_j$, for all $j$.  Note that the Vandermonde matrices (columns and rows) may stop abruptly, immediately after the last coefficient.

Thus, our task for each $q_k$ is to sum the coefficients that match into each column of our various Vandermonde matrices.  We can easily put the first $q_k$ values into a linked list of values, and then cycle back to the start of the list.  Then add the second $q_k$ values to our first values, and cycle back.  In total, this takes at most time proportional to $O(p^3)$.  We do this for a constant number of primes, as we have previously shown, and each addition takes at most time $O(\log{p})$.  Thus the total time for this operation is $O(p^3 \log{p})$, or $O(n \log{n})$.  

The DFT is next calculated for $q_k - 1$ outputs modulo $q_k$, which are the outputs of the Vandermonde matrices modulo each prime $q_k$.  This takes $O \left( q_k \right)^2$ operations for each prime $q_k$, and according to \cite{PrimeSumX}, this sums to 

\begin{align}
	\sum_{\substack{ q_k \text{ Prime} \\ q_k < p}}{ \left( q_k \right)^2 } &= \text{li} (p^{2+1})+ O \left( p^{2+1}e^{-c \sqrt{ \text{ln}(x)}} \right) \\
	                                                                        &= O \left( p^3 / \text{ln}(p^3) \right)
\end{align}

And so we can do this in $O \left( p^3 / \text{ln}(p^3) \right)$ multiplications.  We'll handle this with recursion, soon.  Presently we will say that with a constant number of primes close to $p$, we can easily use the Chinese remainder theorem on our DFTs modulo each $q_k$.  This will take a constant number of multiplications of numbers of size $\log{p}$ bits.  Also we have to cycle through our lists of DFT outputs modulo each $q_k$ more than once, but this won't affect our running time.

We do this fairly easily, as we will see.  Set $\widehat{x_j}(k)$ to be output number $j$ for the DFT done modulo $q_k$.  Then the final result of the entire Polynomial Transform is easily seen to be $\widehat{x_j}$, which is

\begin{align}
	\widehat{x_j} &= \widehat{x_{(j \mod q_1)}}(1) \\
	              &= \widehat{x_{(j \mod q_2)}}(2) \\
							  &= \widehat{x_{(j \mod q_3)}}(3) \\
							  &= \vdots \\
							  &= \widehat{x_{(j \mod q_\alpha)}}(\alpha)
\end{align}

for all $\alpha$ primes.  Here we need to combine the result modulo each prime into the result modulo $q_k$.  To do this, we first observe that each result is done modulo $q_k$ for some $k$, and that we want the final result modulo $p$.  We know that any natural number has a corresponding element modulo $p$, in particular, the number that is equal to $1$ modulo $q_k$ and $0$ modulo every other prime that we use, except $p$.  Then we can simply find what this number is modulo $p$, and multiply our value $\widehat{x_{(j \mod q_k)}}(k)$ by this number.  We do this for all primes $q_k$, add the results, and this is the final result.  The number of primes and multiplications and additions are all constant, and so this takes time that is some constant times the time to add and multiply, or $O(m)$, where $m$ is the time for one multiplication done modulo $p$.  This is done for each and every result, so that the total time for this step is $O(p^3 m)$.

Now we can handle the recursion.  We will say that the initial multiplication has $O(n)$ coefficients, and we can write $n=p^3$, in order to write the recursive multiplications of size $\log{p}$ into the equation.  But this recursion is handled very closely by the iterated logarithm function, yielding time and space $n \left( K^{ \log^*{n} } + \log{n} \right)$, for some undetermined $K$.  This is our final running time.  We note that the space considerations can probably be reduced substantially.


\Section{Correctness}
According to \cite{PolynomialComputations}, for performing the FFT in a ring, for instance modulo  $p$, it suffices to use a \textit{principal} root of unity $\omega$, if $\omega$ has a reciprocal in this ring.  The criterion is that the $n$th principal root of unity $\omega$ is such that:

\begin{align}
	                   \omega^n = 1 \\
\label{Sum Equation} \sum_{i=0}^{n-1}{ \omega^{i j} } = 0, \quad i=1,2,3,\dots,n-1
\end{align}

This, of course, carries over to the DFT and hence the Polynomial transform.  Also, all elements of a finite field are elements of a ring, and so we can surely find a principal root of unity $\omega$ that possesses this property.  Thus, once we find $\omega$, we know that it is a valid $\omega$ value, or principal root of unity, modulo $p$.

The problem now is to prove that the values of the three cube roots of $y$; 1, $\sqrt[3]{y}$, and $\left( \sqrt[3]{y} ]\right)^2$ are correctly calculated.  But this is already done in Section \ref{Algorithm Overview Section} on page \pageref{Algorithm Overview Section}.



\Section{Integer Multiplication}
Here, we can say that we'll transform the multiplication of two $n$ bit numbers into a polynomial transform of size $p^3$.  Now, we know that we'll use transform coefficients of size $\log{p}$, so we can build this right into the multiplication.  So we set, only for this section of the paper, $n = p^3 \log{p}$.  We know from Section \ref{Algorithm Section} that the slowest portions of the Polynomial Transform take time $O(p^3 m)$, where $m$ is the time to multiply.  Since our recursion effectively reduces our coefficients from $n$ to $\log{p}$ in one recursion step, we can easily see that the iterated logarithm will handle this recursion nicely.  Thus it takes time $n k^{ \log^*{n} }$, where $k$ is a constant and $\log^*{n}$ is the iterated logarithm.

\Section{Implications}
A major consequence of sublinearithmic time ($o(n \log{n})$) integer multiplication is that the network coding conjecture\cite{NCC-Paper} is false.

\Section{acknowledgments}
This paper wouldn't be possible without the help of people on StackExchange.com and the math software Mathematica $11.3.0.0$.  In particular, I'd like to thank Jyrki Lahtonen and David Harvey for their help.  I would also like to thank Kenneth Regan of Boston College, and Richard Lipton from Georgia Institute of Technology for their help.

\appendix

\bibliographystyle{plain}
\bibliography{BibFile07000000030}



\end{document}